\documentclass[aps,prl,twocolumn]{revtex4}

\usepackage{graphics}

\begin{document}


\title{Electron capture decay of \\ indium-111 human carbonic anhydrase I: \\
A time differential K X ray coincidence \\ perturbed angular correlation study}


\author{Christopher Haydock}
\email{haydock@mailaps.org}
\altaffiliation[Current address ]{1326 2$^{\mbox{\scriptsize nd}}$ St. NW,
Rochester, MN 55901}
\affiliation{Research Computing Facility \\
Mayo Foundation, Rochester, MN, USA}



\begin{abstract}
The relaxation effects in the
perturbed angular correlation spectra of $^{111}$In
human carbonic anhydrase I (HCA I) are the result of
chemical transmutation and/or the complex Auger cascades that follow the
electron capture decay of $^{111}$In.  Time differential K~X ray
coincidence perturbed angular correlation (PAC) spectroscopy shows that
these relaxation effects are independent of the Auger cascade intensity.
This suggests that chemical transmutation is responsible for the relaxation
effects, and that bond breaking and damage product formation around the decay
site resulting from localized energy deposition by Auger and Coster-Kronig
electrons probably occur in the microsecond time regime.
Numerical simulations of chemical transmutation relaxation effects
in the time differential PAC spectrum of $^{111}$In HCA I are also presented.
\end{abstract}


\maketitle



\section*{Introduction}
Perturbed angular correlation spectroscopy (PAC) is a potential technique
for investigating hot atom chemistry following electron capture decay
of a radionuclide probe~\cite{Adloff78,Boyer84}.  As is the case with
M\"{o}ssbauer emission spectroscopy, it may be possible to identify
the chemical form of the probe nuclide coordination complex
during the nuclear lifetime~\cite{Sano84,Alflen89}.
The average Auger and Coster-Kronig electron yield is significantly
diminished when the vacancy cascade following electron capture includes
the emission of a K X ray as compared to when it does not.
It is in principle possible to study the dependence of hot atom
chemistry on the intensity of Auger and Coster-Kronig emissions
by detecting the nuclear radiations of perturbed angular correlation
or M\"{o}ssbauer emission spectroscopy in coincidence
with K shell X rays~\cite{TI87,Kobayashi79}.  Given the very limited
knowledge of hot atom chemistry of electron capture nuclides that
are possible perturbed angular correlation probes, one may ask only
a yes or no question.  Does or does not the hot atom chemistry
depend electron emission intensity?  To answer this, a chemical
transmutation model and a dose dependent damage model are constructed
for the perturbed angular correlation spectra following electron capture.
Both models encompass several alternative pictures of the hot atom chemistry.
The simplest hot atom chemistry of the chemical transmutation model
is very similar to $\beta^+$ decay on the nanosecond time scale.
Note that $\beta^+$ decay is energetically forbidden for $^{111}$In.
The multiple Auger ionization processes are followed by complete
electron recombination that leaves the daughter nuclide in a metastable state
with both the atomic number and valence decreased by one unit from the parent.
The metastable state is the result of the change in charge rather
that charge neutralization or Auger electron irradiation.
The metastable state of the daughter nuclide decays
within tens of nanoseconds to the ground state coordination complex
of the daughter and ligands.
However, any hot atom chemistry involving an excited state of the daughter
that decays thereafter into the ground state complex is equally consistent
with the chemical transmutation model.  The only requirement is that
the excited state be independent of Auger emission intensity.
The dose dependent damage model pictures radiolytic fragmentation
of probe coordination complex ligands.  The amplitudes
and rates of ligand fluctuations increase with the intensity
of the Auger emissions.
Though uncertain about the details of hot atom chemistry both models
predict perturbed angular correlation spectra.  The comparison of these
spectra with experimental data indicates whether or not the hot atom
chemistry during the nanoseconds following electron capture depends
on Auger emission intensity.

%



\begin{figure}
\resizebox{8.3cm}{!}{\includegraphics{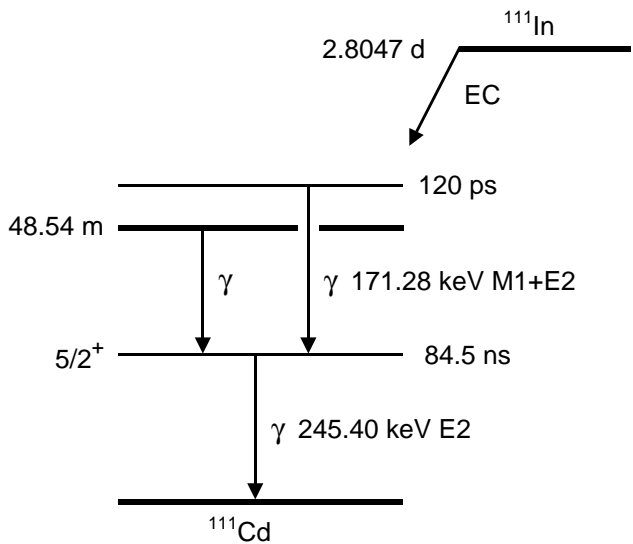}}
\caption{\label{fig:figure1}Partial decay scheme for $^{111}$In and
$^{111m}$Cd~\protect\cite{Shirley78}.  The half-life of each level is labeled.
The 171--245 keV gamma cascade follows the electron capture decay
of $^{111}$In; the 48.5 minute $^{111m}$Cd state decays by the 151--245 keV
gamma cascade.  Both cascades share the same 245 keV $5/2+$
$^{111}$Cd intermediate nuclear state.}
\end{figure}

A partial decay scheme for $^{111}$In and $^{111m}$Cd is shown in Fig.~\ref{fig:figure1}.
Perturbed angular correlation spectra of the isomeric $^{111m}$Cd decay
provide the quadrupole interaction parameters of the ground state
coordination complex in the chemical transmutation model and the ligand
relaxation parameters at zero dose in the dose dependent damage model.
Following $^{111}$In electron capture decay the $^{111}$Cd daughter
is in a 120 picosecond nuclear excited state that initiates the 171--245
keV gamma cascade.  Processes occurring faster than the lifetime
of this initial nuclear state, including dissipation of thermal hot
spots~\cite{Sano84,Kim91}, do not influence perturbed angular
correlation spectra.  The spectra reflect
the hot atom chemistry occurring during the lifetime
of the intermediate state.  Indirect effects involving the diffusion
of water radicals are not observed.  A great variety of radiochemical
and radiobiological experiments show that direct and indirect processes
together result in damage that is proportional to the Auger emission
intensity.  For example, when zinc bovine carbonic anhydrase is X ray
irradiated just above the zinc K edge, enzyme activity and zinc release
assays indicate significant Auger inactivation~\cite{Halpern76}.  The
more intense Auger emissions accompanying $^{111}$In decay can be
expected to inactivate HCA I even more effectively.  The combined
direct and indirect effects are Auger emission intensity dependent.
An Auger dependent effect on the nanosecond time scale can be detected
with K X ray coincidence perturbed angular correlation spectroscopy.

\section*{Materials and Methods}
Human carbonic anhydrase I was isolated from freshly outdated
erythrocytes~\cite{Khalifah77}.  The apoenzyme was obtained by dialysis
at 4$^\circ$C against pyridine-2,6-dicarboxylic acid~\cite{Hunt77}.
HCA I activity was assayed spectrophotometrically with $p$-nitrophenyl
acetate (Sigma Chemical Co., St. Louis, Missouri)
as substrate~\cite{Pocker67}.  Conventional time differential gamma-gamma
perturbed angular correlation spectra of $^{111}$In labeled HCA I were
measured on a movable two detector perturbed angular correlation
spectrometer~\cite{Frauenfelder66}.
A 2 by 2 inch cylindrical BC-404 plastic scintillator
with 0.25 inch centered side bore for internal sample mounting and
RCA 8575 phototube detected the K X rays.  Coincidence gating electronics,
including a time to pulse height converter for coincidence timing K X rays
and the gamma-gamma cascade, selected either K X ray coincidence or
anti-coincidence spectra for storage and analysis.
Carrier free $^{111}$In in 0.05 molar HCl
(Du Pont de Nemours \& Co. Inc., Wilmington, Delaware) was ordered
at two week intervals.  ApoHCA I was incubated for 48 h with
$^{111}$In in 100 mM HEPES buffer (Research Organics Inc., Cleveland, Ohio)
at pH 7.7 with sodium citrate acting as carrier for indium.
In order to hold constant both the volume and total indium
concentration, the specific activity of the $^{111}$In stock was adjusted
with cold indium chloride.  After labeling with indium, the effects of
HCA I rotational motions on the perturbed angular correlation spectrum
were diminished by adding sucrose to a final concentration by weight of 50\%.
Seven labeled protein samples were prepared from each shipment of
$^{111}$In.  A perturbed angular correlation spectrum was accumulated
for 46 h from each sample.  The first and last samples were run
as incubation controls.  These samples were measured for an initial
twelve h as $^{111}$In-citrate; HCA I was added, and the measurement
was continued for another 34 h.  The time integral perturbation
factors accumulated for 20 min intervals monitored the HCA I labeling
reaction.  All spectra from shipments that did not give excellent labeling
were discarded.  Three of the five non-control samples per shipment
were measured in K X ray coincidence mode and two in K X ray anti-coincidence
mode.   A total of 52 coincidence mode samples and 36 anti-coincidence
mode samples were combined for analysis.  All sample preparations and
measurements were done at $21 \pm 1^\circ$C.

Electron capture gamma-gamma perturbed angular correlation spectra
measure the ensemble average electric quadrupole relaxation
of the daughter nucleus spin.  Magnetic dipole interactions are assumed
to be negligible.  The measurable parameters are model dependent.
All models must contain one or more rate constants that characterize
the time dependence of the quadrupole interactions.
In the chemical transmutation model, spin relaxation is the result
of the decay of a metastable state of the daughter nucleus
coordination complex.  This metastable state might be a valence electron,
coordination geometry, or ligand conformation state.  In any case,
the quadrupole interaction parameters of the initial metastable state
are the frequency mean, frequency standard deviation, asymmetry, and the mean
lifetime of the state.  The initial state decays into the ground state
coordination complex.  The interaction parameters of the ground state are
the frequency mean, frequency standard deviation, asymmetry, and Euler
rotation angles relative to the initial state.  The initial and ground
state interaction frequencies are independent Gaussian distributions.
Each individual daughter nucleus interacts with the initial and
ground state coordination complexes at fixed quadrupole frequencies
that are randomly and independently sampled from the respective
Gaussian distributions.  The lifetimes of the initial interactions
are given by an exponential distribution.  In addition to relaxation
from coordination complex decay, the chemical transmutation
model includes relaxation from overall molecular tumbling by
multiplying the spectrum with an exponential factor.

The dose dependent damage model ascribes the spin relaxation to ligand
reorientation in the coordination complex.  The quadrupole interaction
parameters of the coordination complex are the mean frequency,
frequency standard deviation, asymmetry, rotation angle standard deviation,
and jump rate.  These five parameters depend linearly on dose to the
coordination complex and have a baseline value for the zero dose
coordination complex.  The electric field gradient orientation
distribution is defined by rotations from a reference orientation,
where the rotation angle distribution is Gaussian and the rotation axis
distribution is isotropic.  Note that though there is a single reference
orientation in this interaction model, the perturbed angular correlation
spectrum is for an isotropic source because the interactions of each daughter
nucleus are averaged over random orientations when the spectrum
is computed.  The interaction frequency distribution is a Gaussian.
Both the interaction frequency and orientation are sampled after every jump.
The time intervals between jumps are exponentially distributed.

{\setlength{\arraycolsep}{1.2mm}
In both the chemical transmutation and dose dependent damage models,
the interaction of each individual daughter nucleus is piecewise static
and each set of model parameters specifies a probability distribution
function for piecewise static interactions.
The perturbation factor (which is loosely refered to as the
perturbed angular correlation spectrum), is the average over this
probability distribution of the perturbation factor of each piecewise
static interaction.  For a piecewise static interaction averaged over
random source orientations, the perturbation factors are~\cite{Frauenfelder66},
\begin{eqnarray}
G_{kk}(t) & = & \sum_{m_a m_b N} \Big\{ (-1)^{2I+m_a+m_b}
    \nonumber \\ & & \times
  \left( \begin{array}{lll} I & I & k \\ m'_a & -m_a & N \end{array} \right)
  \left( \begin{array}{lll} I & I & k \\ m'_b & -m_b & N \end{array} \right)
    \nonumber \\ & & \times\;
  \langle m_b |\Lambda(t)| m_a \rangle
  \langle m'_b |\Lambda(t)| m'_a \rangle^{\ast} \Big\},
  \label{eq:perturbation}
\end{eqnarray}
where $k$ is an even index, $t$ is the delay time between emission
of the two gamma rays from the daughter nucleus gamma-gamma cascade,
$I$ is the spin of the intermediate nuclear state,
$m_a$, $m'_a$, $m_b$, and $m'_b$ index magnetic substates of
the intermediate state, the matrix element coefficients are the
Wigner 3-$j$ symbols, and $\Lambda(t)$ is the piecewise static
time evolution operator.  Since only $G_{22}(t)$ is usually measurable,
it is refered to as the perturbation factor.
The index $N$ takes all integer values with absolute
value less than or equal to $k$ and the primed magnetic substate indices
are given by $m'_a  -m_a + N = 0$ and $m'_b  -m_b + N = 0$.
Since the perturbation factor is always computed at a series
of time points, it is appropriate to express the piecewise
static time evolution operator as a series operator product
of interval evolution operators.  The interval evolution operator
for the time interval $(t',t)$ is a series operator product
of static evolution operators,
\begin{equation}
\Lambda(t',t) = \prod_{j=J'}^J e^{-iH_j t_j/\hbar},
  \label{eq:evolution}
\end{equation}
where $H_j$ is the $j^{\mbox{\scriptsize th}}$ time independent interaction
Hamiltonian, $J$ is the index of the first Hamiltonian with active interval
overlapping $(t',t)$, $J'$ is the index of the last Hamiltonian with active
interval overlapping $(t',t)$, and $t_j$ is the time overlap of $(t',t)$ and
the active time interval of the $j^{\mbox{\scriptsize th}}$ Hamiltonian.
Matthias et al.~\cite{Matthais63} give the electric quadrupole Hamiltonian
matrix elements for arbitrary intermediate spin, interaction frequency,
asymmetry parameter, and $y$ convention \cite{Goldstein80} orientation.
The quadrupole interaction frequency is specified by the angular
frequency~$\omega_0$.  By definition, $\omega_0$ equals 3 times
the quadrupole interaction frequency for integer spin and 6 times
the quadrupole interaction frequency for half integer spin.
When the electric quadrupole Hamiltonian is axially symmetric
$\omega_0 \hbar$ equals the smallest nonvanishing eigenvalue difference.
The angular frequency $\omega_0$ is frequently reported with correct
numerical value but with the erroneous dimension of Hertz~\cite{Bauer85}.
}

The piecewise static perturbation factors were evaluated
by repeated extension of the time evolution operators
in increments of $0.025(2\pi/\omega_0)$, where $\omega_0$ was the mean
angular frequency of the ground state or zero dose coordination complex.
The evolution operator at each time point was extended to the
next time point by matrix multiplication with a time interval evolution
operator.  Each time interval evolution operator was
a product of static evolution operators with the Hamiltonian indices
and overlap intervals indicated in Eq.~(\ref{eq:evolution}).
The interaction Hamiltonians were diagonalized~\cite{EISPACK76};
the static evolution operators were expressed in diagonal form with
the eigenvalues, transformed back into the magnetic substate
representation with the eigenvectors, and multiplied together to
give the time interval evolution operators.
Since in the parameter domain of interest the perturbation factor time
intervals were an order of magnitude smaller than the typical static
Hamiltonian active time interval, most interval evolution operators
were a single static evolution operator and each Hamiltonian
diagonalization generated many interval evolution operators.
The time interval evolution operators were repeatedly multiplied to give
the evolution operators at a series of 160 time points.
The matrix elements of the evolution operators were inserted in
Eq.~(\ref{eq:perturbation}) to give the piecewise static
perturbation factors at these time points.

The piecewise static perturbation factor probability distribution
function was averaged by Monte Carlo integration~\cite{James80}.
Perturbation factor error estimates were calculated from
the deviation of subaverages around the overall average as a function
of the spectrum delay time.
Gaussian interaction frequency and rotation angle distributions
were generated by summing uniform random variates.
Excited state and orientation lifetimes were generated by
logarithmic transform of uniform random numbers on the unit interval.
Isotropic rotation directions were generated by an acceptance-rejection
technique~\cite{Allen_Tildesley89}.
Random numbers uniform on the unit interval were generated by
a 32-bit linear congruential pseudo-random generator with divisor $2^{31}-1$
and multiplier~$7^5$.

Initially the qualitative dependence of the perturbation factor
on parameter space of the chemical transmutation and the
dose dependent damage models was coarsely surveyed.
The chemical transmutation model metastable state lifetime,
ground state rotation angle, and relative magnitude of the metastable
and ground state mean interaction frequency parameters were varied.
Based on the preliminary survey and known ground state coordination
complex parameters, a crude estimate was made of the parameter
point that fit the data.  This fit was refined
by varying the unknown parameters one
at a time until the residuals were about the same magnitude
as the experimental error.  The chemical transmutation model
ground state frequency mean, frequency standard deviation,
and asymmetry and the overall tumbling relaxation time constant
were fixed at the known values.  The Euler rotation angles relative
to the initial state were fixed at values found in the preliminary survey
to give strong spin relaxation.  The initial metastable state was
assumed to be axially symmetric, and the frequency mean and frequency
standard deviation were varied.

The preliminary dose dependent damage model parameter survey
fixed the mean frequency, frequency standard deviation, and asymmetry
parameters at the known zero dose coordination complex values, and varied
the rotation angle standard deviation and jump rate parameters.
By comparing of the results of this search and the perturbation factor
measured by Bauer et al.~\cite{Bauer76} with the assumption that spin
relaxation is due to molecular tumbling, values were identified for the
rotation angle standard deviation and jump rate parameters approximately
equivalent to the molecular tumbling time constant of the zero
dose coordination complex.  The preliminary search also yielded
the initial parameter point for fitting the dose dependent damage
model to the data.  The mean frequency, frequency standard deviation,
asymmetry, rotation angle standard deviation, and jump rate parameters
were all varied.

During the preliminary parameter space searches and parameter refinement,
the perturbation factors were computed by averaging $10^3$ piecewise
static interactions.  The standard deviations of these perturbation
factors were fairly independent of spectrum delay time and
typically about 0.01.  The displayed perturbation factors were
averaged over $10^4$ interactions to give a standard deviation of
about 0.003.  All calculations were done on a DEC VAX 3600
or Silicon Graphics 4D workstation.

\section*{Results and Discussion}
\begin{figure}
\resizebox{8.3cm}{!}{\includegraphics{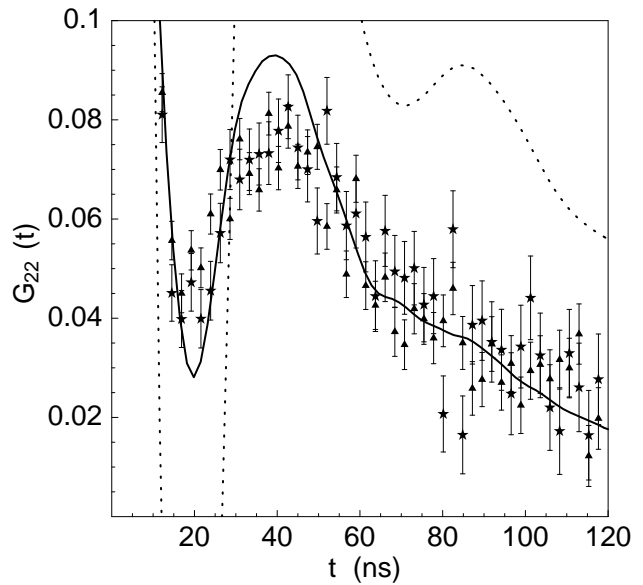}}
\caption{\label{fig:figure2}Chemical transmutation model for HCA I perturbed
angular correlation spectra.  The triangles with standard error markers are
the K X ray anti-coincidence spectrum and the stars with standard error
markers are the K X ray coincidence spectrum.
The solid curve is an approximate fit of the chemical transmutation
model to the data.  The dotted curve is the least squares fit
of Bauer et al.~\protect\cite{Bauer76} to the $^{111m}$Cd HCA I perturbed
angular correlation spectrum.}
\end{figure}

\begin{figure}
\resizebox{8.3cm}{!}{\includegraphics{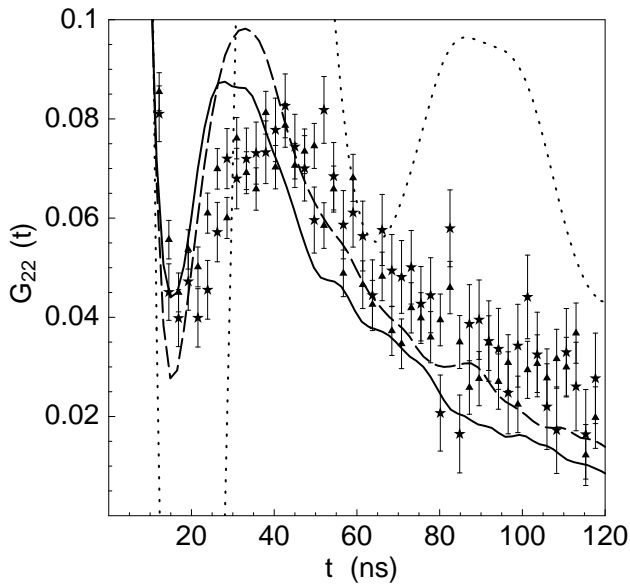}}
\caption{\label{fig:figure3}Dose dependent damage model for HCA I perturbed
angular correlation spectra.  The experimental data is identical to that
displayed in Fig.~2.  The solid curve is an approximate fit of
the dose dependent damage model to the data.  The dotted curve
is an approximate fit to the $^{111m}$Cd HCA I perturbed angular
correlation spectrum with relaxation from ligand reorientation
rather than molecular tumbling.  The dashed curve is the
dose dependent damage model for the K X ray coincidence HCA I
perturbed angular correlation spectrum derived by linear interpolation
of the model parameters for the $^{111m}$Cd and
K X ray anti-coincidence spectra.}
\end{figure}

The measured K X ray coincidence and anti-coincidence perturbed
angular correlation spectra for $^{111}$In HCA I in 50\% sucrose
are shown in both Figs.\ \ref{fig:figure2} \&~\ref{fig:figure3}.
Figure \ref{fig:figure2} shows theoretical spectra for the chemical
transmutation model at parameter values listed in Table~\ref{tab:table1}
and Fig.~\ref{fig:figure3} shows theoretical spectra for the dose
dependent damage model at parameter values listed in Table~\ref{tab:table2}.
The procedure for approximately fitting the models is given in the
methods section.
Relaxation in the chemical transmutation model spectra results from
metastable coordination complex decay and molecular tumbling.
Evidently these mechanisms can account for all relaxation experimentally
observed in $^{111}$In HCA I perturbed angular correlation spectra.
Since the coordination complex initial metastable state is assumed to be
independent of the Auger emission intensity, the chemical transmutation model
predicts equality of the K X ray coincidence and anti-coincidence spectra.

%


\begin{table}
\caption{\label{tab:table1}Chemical transmutation model quadrupole interaction
parameters for perturbed angular correlation spectra of HCA I.}
\begin{ruledtabular}
\begin{tabular}{lll}
   &$ ^{111m}$Cd & $^{111}$In \\ \hline
Metastable state && \\
$\;\;$ Freq. mean (s$^{-1}$)&   -   &  $133 \times 10^6$ \\
$\;\;$ Freq. SD             &   -   &  50\% \\
$\;\;$ Asymmetry                &   -   &  0.0  \\
$\;\;$ Lifetime (ns)            &  0.0  &  66   \\
Ground state && \\
$\;\;$ Freq. mean (s$^{-1}$)& $95 \times 10^6$
                                            & $95 \times 10^6$ \\
$\;\;$ Freq. SD                     &  18\% &  18\% \\
$\;\;$ Asymmetry                        &  0.68 &  0.68 \\
$\;\;$ Euler angles ($\phi,\theta,\psi$) &   -   & $0,\pi/2,0$ \\
 Tumbling lifetime(ns)                      &  81   &  81 \\
\end{tabular}
\end{ruledtabular}
\end{table}

\begin{table}
\caption{\label{tab:table2}Dose dependent damage model quadrupole
interaction parameters.}
\begin{ruledtabular}
\begin{tabular}{llll}
   & $^{111m}$Cd & $^{111}$In ($k+$)
                        & $^{111}$In ($k-$) \\  \hline
Freq. mean (s$^{-1}$)  &  $95 \times 10^6$
                           & $127 \times 10^6$ & $133 \times 10^6$ \\
Freq. SD               &  18\% & 47.6\%  &  53\% \\
Asymmetry                  &  0.68 & 0.468   &  0.43 \\
Rotation SD                &  25$^\circ$ & 37.7$^\circ$ &  40$^\circ$ \\
Jump lifetime (ns)         &  26.5 & 18.1    &  16.5 \\
\end{tabular}
\end{ruledtabular}
\end{table}

Relaxation in the dose dependent damage model spectra results from
ligand reorientation.  The absence of molecular tumbling relaxation
in this model is a matter of convenience.  Additional high precision
$^{111m}$Cd HCAB I spectra would be required to establish the
correct balance of relaxation due to internal and overall motion.
The theoretical K X ray coincidence and anti-coincidence spectra
for the dose dependent damage model depend upon the relative Auger
emission intensities.  The electron capture decay of $^{111}$In
in coincidence with a cadmium K X ray yields on average 11.0 Auger
and Coster-Kronig electrons and in anti-coincidence yields on average
16.8 electrons (R. W. Howell, private communication).  The anti-coincidence
yield must be corrected for the efficiency of detection
for K X rays.  The corrected anti-coincidence yield,
\begin{equation}
Y'_{k-} =
  \frac{(1-\omega)}{1-\epsilon\omega}Y_{k-}
 +\frac{\omega(1-\epsilon)}{1-\epsilon\omega}Y_{k+},
  \label{eq:correction}
\end{equation}
where $\epsilon$ is the K X ray detection efficiency, $\omega$ is the
probability per decay of K X ray emission, $Y_{k-}$
is the perfect anti-coincidence yield, and $Y_{k+}$
the K X ray coincidence yield.  At the estimated detection efficiency
for cadmium K X rays of 23\%, the corrected anti-coincidence Auger
electron yield is 13.0.
The ratio $Y_{k+}/Y'_{k-} = 11/13$
interpolates the K X ray coincidence parameters between the zero dose
and K X ray anti-coincidence parameters in the dose dependent damage model.
The ratio of the difference between the K X ray coincidence and zero dose
parameters to the difference between the K X ray anti-coincidence
and zero dose parameters is 11/13, see Table \ref{tab:table2}.
The experimental $^{111}$In HCA I K X ray coincidence and anti-coincidence
spectra are identical.  However, since the difference in the theoretical
K X ray coincidence and anti-coincidence spectra is small enough
relative to the experimental error, the dose dependent damage model
may not be confidently rejected.


\section*{Conclusions}
The indistinguishability of the experimental $^{111}$In HCA I
K X ray coincidence and anti-coincidence spectra suggests
that during the first $10^{-8}$ seconds following electron capture
decay the hot atom chemistry is independent of the Auger and
Coster-Kronig electron emission intensity.  As demonstrated by
the dose dependent damage model, the experimental data do not
conclusively rule out dependence of hot atom chemistry on emission intensity.
The difference between the HCA I K X ray coincidence and anti-coincidence
spectra could be significantly increased by detecting K$_{\beta}$
X rays for the coincidence spectrum, increasing the detection efficiency
of all K X rays for the anti-coincidence spectrum, and possibly
by measuring at liquid nitrogen temperature.



\begin{acknowledgments}
Special thanks are due to Kandula S. R. Sastry and Franklyn G. Prendergast
for their continued encouragement and support of this project.
Lynda McDowell purified HCA I and helped with indium binding assays.
Heiner Winkler introduced me to the Monte Carlo procedure~\cite{Gerdau76}
for computing a perturbation factor by averaging over the probability
distribution of piecewise static interactions.
Roger Howell provided electron yields.
Alexander Halpern has called to my attention
references~\cite{Adloff78,Boyer84,Sano84,Halpern76}.
This work was supported in part by National Institutes of Health
grant 1 R03 RR02219-01.
\end{acknowledgments}

\bibliography{auger}

\begin{thebibliography}{21}
\expandafter\ifx\csname natexlab\endcsname\relax\def\natexlab#1{#1}\fi
\expandafter\ifx\csname bibnamefont\endcsname\relax
  \def\bibnamefont#1{#1}\fi
\expandafter\ifx\csname bibfnamefont\endcsname\relax
  \def\bibfnamefont#1{#1}\fi
\expandafter\ifx\csname citenamefont\endcsname\relax
  \def\citenamefont#1{#1}\fi
\expandafter\ifx\csname url\endcsname\relax
  \def\url#1{\texttt{#1}}\fi
\expandafter\ifx\csname urlprefix\endcsname\relax\def\urlprefix{URL }\fi
\providecommand{\bibinfo}[2]{#2}
\providecommand{\eprint}[2][]{\url{#2}}

\bibitem[{\citenamefont{Adloff}(1978)}]{Adloff78}
\bibinfo{author}{\bibfnamefont{J.~P.} \bibnamefont{Adloff}},
  \bibinfo{journal}{Radiochim. Acta} \textbf{\bibinfo{volume}{25}},
  \bibinfo{pages}{57} (\bibinfo{year}{1978}).

\bibitem[{\citenamefont{Boyer and Baudry}(1984)}]{Boyer84}
\bibinfo{author}{\bibfnamefont{P.}~\bibnamefont{Boyer}} \bibnamefont{and}
  \bibinfo{author}{\bibfnamefont{A.}~\bibnamefont{Baudry}}, in
  \emph{\bibinfo{booktitle}{Hot Atom Chemistry}}, edited by
  \bibinfo{editor}{\bibfnamefont{T.}~\bibnamefont{Matsuura}}
  (\bibinfo{publisher}{Elsevier}, \bibinfo{address}{Amsterdam},
  \bibinfo{year}{1984}), pp. \bibinfo{pages}{315--347}.

\bibitem[{\citenamefont{Sano and G\"{u}tlich}(1984)}]{Sano84}
\bibinfo{author}{\bibfnamefont{H.}~\bibnamefont{Sano}} \bibnamefont{and}
  \bibinfo{author}{\bibfnamefont{P.}~\bibnamefont{G\"{u}tlich}}, in
  \emph{\bibinfo{booktitle}{Hot Atom Chemistry}}, edited by
  \bibinfo{editor}{\bibfnamefont{T.}~\bibnamefont{Matsuura}}
  (\bibinfo{publisher}{Elsevier}, \bibinfo{address}{Amsterdam},
  \bibinfo{year}{1984}), pp. \bibinfo{pages}{265--302}.

\bibitem[{\citenamefont{Alflen et~al.}(1989)\citenamefont{Alflen, Hennen,
  Tuczek, Spiering, G\"{u}tlich, and Kajcsos}}]{Alflen89}
\bibinfo{author}{\bibfnamefont{M.}~\bibnamefont{Alflen}},
  \bibinfo{author}{\bibfnamefont{C.}~\bibnamefont{Hennen}},
  \bibinfo{author}{\bibfnamefont{F.}~\bibnamefont{Tuczek}},
  \bibinfo{author}{\bibfnamefont{H.}~\bibnamefont{Spiering}},
  \bibinfo{author}{\bibfnamefont{P.}~\bibnamefont{G\"{u}tlich}},
  \bibnamefont{and} \bibinfo{author}{\bibfnamefont{Z.}~\bibnamefont{Kajcsos}},
  \bibinfo{journal}{Hyperfine Interact.} \textbf{\bibinfo{volume}{47}},
  \bibinfo{pages}{115} (\bibinfo{year}{1989}).

\bibitem[{\citenamefont{Haydock and Sastry}(1987)}]{TI87}
\bibinfo{author}{\bibfnamefont{C.}~\bibnamefont{Haydock}} \bibnamefont{and}
  \bibinfo{author}{\bibfnamefont{K.~S.~R.} \bibnamefont{Sastry}}, in
  \emph{\bibinfo{booktitle}{Proceedings of the 8th International Congress of
  Radiation Research}} (\bibinfo{year}{1987}), vol.~\bibinfo{volume}{1},
  p.~\bibinfo{pages}{67}.

\bibitem[{\citenamefont{Kobayashi et~al.}(1979)\citenamefont{Kobayashi,
  Fukumura, Kitahara, and Shimizu}}]{Kobayashi79}
\bibinfo{author}{\bibfnamefont{T.}~\bibnamefont{Kobayashi}},
  \bibinfo{author}{\bibfnamefont{K.}~\bibnamefont{Fukumura}},
  \bibinfo{author}{\bibfnamefont{T.}~\bibnamefont{Kitahara}}, \bibnamefont{and}
  \bibinfo{author}{\bibfnamefont{S.}~\bibnamefont{Shimizu}},
  \bibinfo{journal}{J. Phys. Colloq.} pp. \bibinfo{pages}{C28--29}
  (\bibinfo{year}{1979}).

\bibitem[{\citenamefont{Lederer and Shirley}(1978)}]{Shirley78}
\bibinfo{editor}{\bibfnamefont{C.~M.} \bibnamefont{Lederer}} \bibnamefont{and}
  \bibinfo{editor}{\bibfnamefont{V.~S.} \bibnamefont{Shirley}}, eds.,
  \emph{\bibinfo{title}{Table of Isotopes}} (\bibinfo{publisher}{John Wiley},
  \bibinfo{address}{New York}, \bibinfo{year}{1978}), pp.
  \bibinfo{pages}{516--523}, \bibinfo{edition}{seventh} ed.,
  \bibinfo{note}{$\mbox{A}=111$},
  \urlprefix\url{http://ie.lbl.gov/toi/nucSearch.asp}.

\bibitem[{\citenamefont{Kim and Dlott}(1991)}]{Kim91}
\bibinfo{author}{\bibfnamefont{H.}~\bibnamefont{Kim}} \bibnamefont{and}
  \bibinfo{author}{\bibfnamefont{D.~D.} \bibnamefont{Dlott}},
  \bibinfo{journal}{J. Chem. Phys.} \textbf{\bibinfo{volume}{94}},
  \bibinfo{pages}{8203} (\bibinfo{year}{1991}).

\bibitem[{\citenamefont{Diehn et~al.}(1976)\citenamefont{Diehn, Halpern, and
  St\"{o}cklin}}]{Halpern76}
\bibinfo{author}{\bibfnamefont{B.}~\bibnamefont{Diehn}},
  \bibinfo{author}{\bibfnamefont{A.}~\bibnamefont{Halpern}}, \bibnamefont{and}
  \bibinfo{author}{\bibfnamefont{G.}~\bibnamefont{St\"{o}cklin}},
  \bibinfo{journal}{J. Am. Chem. Soc.} \textbf{\bibinfo{volume}{98}},
  \bibinfo{pages}{1077} (\bibinfo{year}{1976}).

\bibitem[{\citenamefont{Khalifah et~al.}(1977)\citenamefont{Khalifah, Strader,
  Bryant, and Gibson}}]{Khalifah77}
\bibinfo{author}{\bibfnamefont{R.~G.} \bibnamefont{Khalifah}},
  \bibinfo{author}{\bibfnamefont{D.~J.} \bibnamefont{Strader}},
  \bibinfo{author}{\bibfnamefont{S.~H.} \bibnamefont{Bryant}},
  \bibnamefont{and} \bibinfo{author}{\bibfnamefont{S.~M.}
  \bibnamefont{Gibson}}, \bibinfo{journal}{Biochemistry}
  \textbf{\bibinfo{volume}{16}}, \bibinfo{pages}{2241} (\bibinfo{year}{1977}).

\bibitem[{\citenamefont{Hunt et~al.}(1977)\citenamefont{Hunt, Rhee, and
  Storm}}]{Hunt77}
\bibinfo{author}{\bibfnamefont{J.~B.} \bibnamefont{Hunt}},
  \bibinfo{author}{\bibfnamefont{M.-J.} \bibnamefont{Rhee}}, \bibnamefont{and}
  \bibinfo{author}{\bibfnamefont{C.~B.} \bibnamefont{Storm}},
  \bibinfo{journal}{Anal. Biochem.} \textbf{\bibinfo{volume}{79}},
  \bibinfo{pages}{614} (\bibinfo{year}{1977}).

\bibitem[{\citenamefont{Pocker and Stone}(1967)}]{Pocker67}
\bibinfo{author}{\bibfnamefont{Y.}~\bibnamefont{Pocker}} \bibnamefont{and}
  \bibinfo{author}{\bibfnamefont{J.~T.} \bibnamefont{Stone}},
  \bibinfo{journal}{Biochemistry} \textbf{\bibinfo{volume}{6}},
  \bibinfo{pages}{668} (\bibinfo{year}{1967}).

\bibitem[{\citenamefont{Frauenfelder and Steffen}(1965)}]{Frauenfelder66}
\bibinfo{author}{\bibfnamefont{H.}~\bibnamefont{Frauenfelder}}
  \bibnamefont{and} \bibinfo{author}{\bibfnamefont{R.~M.}
  \bibnamefont{Steffen}}, in \emph{\bibinfo{booktitle}{Alpha-, Beta- and Gamma-
  Ray Spectroscopy}}, edited by
  \bibinfo{editor}{\bibfnamefont{K.}~\bibnamefont{Siegbahn}}
  (\bibinfo{publisher}{North-Holland}, \bibinfo{address}{Amsterdam},
  \bibinfo{year}{1965}), vol.~\bibinfo{volume}{2}, chap.
  \bibinfo{chapter}{XIXA}, pp. \bibinfo{pages}{997--1198}.

\bibitem[{\citenamefont{Matthais et~al.}(1963)\citenamefont{Matthais,
  Schneider, and Steffen}}]{Matthais63}
\bibinfo{author}{\bibfnamefont{E.}~\bibnamefont{Matthais}},
  \bibinfo{author}{\bibfnamefont{W.}~\bibnamefont{Schneider}},
  \bibnamefont{and} \bibinfo{author}{\bibfnamefont{R.~M.}
  \bibnamefont{Steffen}}, \bibinfo{journal}{Ark. Fys.}
  \textbf{\bibinfo{volume}{24}}, \bibinfo{pages}{97} (\bibinfo{year}{1963}).

\bibitem[{\citenamefont{Goldstein}(1980)}]{Goldstein80}
\bibinfo{author}{\bibfnamefont{H.}~\bibnamefont{Goldstein}},
  \emph{\bibinfo{title}{Classical Mechanics}}
  (\bibinfo{publisher}{Addison-Wesley}, \bibinfo{address}{Massachusetts},
  \bibinfo{year}{1980}), pp. \bibinfo{pages}{606--610}, \bibinfo{edition}{2nd}
  ed., \bibinfo{note}{appendix B}.

\bibitem[{\citenamefont{Bauer}(1985)}]{Bauer85}
\bibinfo{author}{\bibfnamefont{R.}~\bibnamefont{Bauer}}, \bibinfo{journal}{Q.
  Rev. Biophys.} \textbf{\bibinfo{volume}{18}}, \bibinfo{pages}{1}
  (\bibinfo{year}{1985}).

\bibitem[{\citenamefont{Smith et~al.}(1976)\citenamefont{Smith, Boyle,
  Dongarra, Garbow, Ikebe, Klema, and Moler}}]{EISPACK76}
\bibinfo{author}{\bibfnamefont{B.~T.} \bibnamefont{Smith}},
  \bibinfo{author}{\bibfnamefont{J.~M.} \bibnamefont{Boyle}},
  \bibinfo{author}{\bibfnamefont{J.~J.} \bibnamefont{Dongarra}},
  \bibinfo{author}{\bibfnamefont{B.~S.} \bibnamefont{Garbow}},
  \bibinfo{author}{\bibfnamefont{Y.}~\bibnamefont{Ikebe}},
  \bibinfo{author}{\bibfnamefont{V.~C.} \bibnamefont{Klema}}, \bibnamefont{and}
  \bibinfo{author}{\bibfnamefont{C.~B.} \bibnamefont{Moler}},
  \emph{\bibinfo{title}{Matrix Eigensystem Routines - EISPACK Guide}}
  (\bibinfo{publisher}{Springer-Verlag}, \bibinfo{address}{Berlin},
  \bibinfo{year}{1976}), no.~\bibinfo{number}{6} in \bibinfo{series}{Lecture
  Notes in Computer Science}, \bibinfo{edition}{2nd} ed.

\bibitem[{\citenamefont{James}(1980)}]{James80}
\bibinfo{author}{\bibfnamefont{F.}~\bibnamefont{James}}, \bibinfo{journal}{Rep.
  Prog. Phys.} \textbf{\bibinfo{volume}{43}}, \bibinfo{pages}{1145}
  (\bibinfo{year}{1980}).

\bibitem[{\citenamefont{Allen and Tildesley}(1989)}]{Allen_Tildesley89}
\bibinfo{author}{\bibfnamefont{M.~P.} \bibnamefont{Allen}} \bibnamefont{and}
  \bibinfo{author}{\bibfnamefont{D.~J.} \bibnamefont{Tildesley}},
  \emph{\bibinfo{title}{Computer Simulation of Liquids}}
  (\bibinfo{publisher}{Clarendon Press}, \bibinfo{address}{Oxford},
  \bibinfo{year}{1989}), pp. \bibinfo{pages}{345--351}, \bibinfo{note}{appendix
  G}.

\bibitem[{\citenamefont{Bauer et~al.}(1976)\citenamefont{Bauer, Limkilde, and
  Johansen}}]{Bauer76}
\bibinfo{author}{\bibfnamefont{R.}~\bibnamefont{Bauer}},
  \bibinfo{author}{\bibfnamefont{P.}~\bibnamefont{Limkilde}}, \bibnamefont{and}
  \bibinfo{author}{\bibfnamefont{J.~T.} \bibnamefont{Johansen}},
  \bibinfo{journal}{Biochemistry} \textbf{\bibinfo{volume}{15}},
  \bibinfo{pages}{334} (\bibinfo{year}{1976}).

\bibitem[{\citenamefont{Gerdau et~al.}(1976)\citenamefont{Gerdau, Winkler,
  Giese, Gebert, and Braunsfurth}}]{Gerdau76}
\bibinfo{author}{\bibfnamefont{E.}~\bibnamefont{Gerdau}},
  \bibinfo{author}{\bibfnamefont{H.}~\bibnamefont{Winkler}},
  \bibinfo{author}{\bibfnamefont{B.}~\bibnamefont{Giese}},
  \bibinfo{author}{\bibfnamefont{W.}~\bibnamefont{Gebert}}, \bibnamefont{and}
  \bibinfo{author}{\bibfnamefont{J.}~\bibnamefont{Braunsfurth}},
  \bibinfo{journal}{Hyperfine Interact.} \textbf{\bibinfo{volume}{1}},
  \bibinfo{pages}{469} (\bibinfo{year}{1976}).

\end{thebibliography}

\end{document}